*********************************************************************
\documentstyle[aps,epsf,twocolumn]{revtex}

\newcommand{\be}{\begin{eqnarray}}
\newcommand{\ee}{\end{eqnarray}}

\begin{document}

\draft
\title{\bf  A Master Formula Approach to Chiral Symmetry Breaking}

\author{{\bf Hidenaga Yamagishi}$^1$
and {\bf Ismail Zahed}$^2$}

\address{$^1$4 Chome 11-16-502, Shimomeguro, Meguro, Tokyo, Japan. 153;\\
$^2$Department of Physics, SUNY, Stony Brook, New York 11794, USA.}
\date{\today}
\maketitle

\begin{abstract}
We find that various results of current algebra at tree level and beyond
can be directly obtained from a master formula, without use of chiral
perturbation theory or effective Lagrangians.
Application is made to $\pi \pi$ scattering, where it is shown
that the bulk of the $\rho$ contribution can be determined in a model
independent way.
\end{abstract}
\narrowtext

Consider an action whose kinetic part is invariant under chiral
$SU_L(2)\times SU_R(2)$ with a scalar-isoscalar mass term in the $(2,2)$
representation. Examples are two flavor QCD or
sigma models. The symmetry properties of the theory
may be expressed by gauging the kinetic part with c-number external
fields $v_{\mu}^a$ and $a_{\mu}^a$, and extending the mass term to include
couplings with scalar and pseudoscalar fields $s$ and $p^a$. For two-flavor
QCD, the relevant part of the action reads
\be
{\bf I} =&&+ \int d^4x \overline{q}\gamma^{\mu}
\bigg(i\partial_{\mu} + G_{\mu} + v_{\mu}^a\frac {\tau^a}2 +
a_{\mu}^a\frac {\tau^a}2\gamma_5 \bigg) q \nonumber\\&&
-\frac {\hat{m}}{m_{\pi}^2}
\int d^4x \overline{q}\bigg( m_{\pi}^2 + s -i\gamma_5 \tau^a p^a \bigg) q
\label{A1}
\ee
where $m_{\pi}$ is the pion mass.
We will assume that $\phi = (v_{\mu}^a, a_{\mu}^a, s, p^a)$ are
smooth functions that fall off rapidly at infinity.

Currents and  densities
${\cal O} = ({\bf V}, {\bf A}, f_{\pi}\sigma , f_{\pi} \pi )$
may be introduced as
\be
{\cal O} (x) = \frac{\delta {\bf I}}{\delta \phi (x)}
\label{A2}
\ee
which obey the Veltman-Bell equations \cite{veltman}
\be
\nabla^{\mu}{\bf V}_{\mu} +\b a^{\mu} {\bf A}_{\mu} + f_{\pi} {\b p}\,\pi = 0
\label{A3}
\ee
\be
\nabla^{\mu}{\bf A}_{\mu} +\b a^{\mu} {\bf V}_{\mu} -f_{\pi} (m_{\pi}^2 +s )
\,\pi + f_{\pi} p\, \sigma = 0
\label{A4}
\ee
where $\nabla_{\mu} =\partial_{\mu }\bf 1 +\b v_{\mu}$ is the vector covariant
derivative, ${\b a}_{\mu}^{ac} = \epsilon^{abc} a_{\mu}^b$,
${\b p}^{ac} = \epsilon^{abc} p^b$, and $f_{\pi}$
is the pion decay constant. In the above,
we have used the fact that the Bardeen anomaly
\cite{bardeen} and the Wess-Zumino term \cite{wess} vanish
for $SU_L(2)\times SU_R(2)$.
Introducing the extended S-matrix ${\cal S}$, holding the incoming fields
fixed, and using the Schwinger action principle \cite{schwinger}
imply
\be
<\beta \,\,{\rm in} |\delta{\cal S} |\alpha \,\,{\rm in} > = i
<\beta \,\,{\rm in} |{\cal S}\delta {\bf I} |\alpha \,\,{\rm in} > \,\,.
\label{A5}
\ee
This result together with asymptotic completeness,
yield the Peierls-Dyson formula \cite{peirls}
\be
{\cal O} (x) = -i{\cal S}^{\dagger} \frac{\delta {\cal S}}{\delta \phi (x)}
\,\,\,.
\label{A6}
\ee

It follows from the Veltman-Bell equations (\ref{A3}-\ref{A4}) that
\be
&&\bigg(\nabla_{\mu}^{ac}\frac{\delta}{\delta v_{\mu}^c (x)} +
                  {\b a}_{\mu}^{ac} (x) \frac{\delta}{\delta a^c_{\mu }(x)}
+{\b p}^{ac} (x) \frac{\delta}{\delta p^c (x)}\bigg) {\cal S} = \nonumber\\&&
\bigg( {\bf X}^a_V(x) +{\b p}^{ac} (x) \frac {\delta}{\delta p^c (x)}
\bigg) {\cal S} = 0
\label{A7}
\ee
\be
&&\bigg(\nabla_{\mu}^{ac}\frac{\delta}{\delta a_{\mu}^c (x)} +
                  {\b a}_{\mu}^{ac} (x) \frac{\delta}{\delta v^c_{\mu }(x)}
\nonumber\\&&
-(m_{\pi}^2 +s(x) ) \frac{\delta}{\delta p^a (x)}
+{ p^a (x)} \frac {\delta}{\delta s (x)} \bigg) {\cal S} = \nonumber\\&&
\bigg( {\bf X}^a_A (x) -(m_{\pi}^2 +s (x)) \frac{\delta}{\delta p^a (x)}
+{ p}^a (x) \frac {\delta}{\delta s (x)} \bigg) {\cal S} = 0
\nonumber\\
\label{A8}
\ee
where ${\bf X}_V$ and ${\bf X}_A$
are the generators of local $SU_L(2)\times SU_R(2)$.

We further require
\be
<0| {\bf A}_{\mu}^a (x) |\pi^b (p) > = if_{\pi} \delta^{ab} p_{\mu}
\,\,e^{-ip\cdot x}\,\,\,.
\label{A11}
\ee
In the absence of stable axial vector or other pseudoscalar mesons, this is
equivalent to the asymptotic conditions ($x^0\rightarrow\mp\infty$)
\be
{\bf A}_{\mu}^a (x)\rightarrow -f_{\pi} \partial_{\mu} \pi_{\rm in, out}^a (x)
\nonumber
\ee
and
\be
\partial^{\mu}{\bf A}_{\mu}^a (x)\rightarrow +f_{\pi} m_{\pi}^2
\pi_{\rm in, out}^a (x)
\label{A12}
\ee
where $\pi_{\rm in}$ and $\pi_{\rm out}$ are free incoming and outgoing
pion fields. Comparison of (\ref{A12}) with (\ref{A4}) shows that $\pi$ is a
normalized interpolating field.

To incorporate (\ref{A12}) into (\ref{A7}-\ref{A8}) we introduce a modified
action
\be
\hat{\bf I} = {\bf I} -f_{\pi}^2 \int d^4x \bigg( s (x) +
\frac 12 a^{\mu} (x) \cdot a_{\mu} (x) \bigg)\,\,\,,
\label{A13}
\ee
the corresponding  extended S-matrix
\be
\hat{\cal S} = {\cal S}\,\, {\rm exp}\bigg({-if_{\pi}^2\int d^4x
\bigg( s (x) + \frac 12 a^{\mu} (x) \cdot a_{\mu} (x) \bigg)}\bigg)\,\,\,,
\label{A14}
\ee
and a change of variable $p=J/f_{\pi}-\nabla^{\mu} a_{\mu}$.
Taking $\hat\phi =(v_{\mu}^a , a_{\mu}^a, s, J^a )$ as independent
variables, modified currents and densities
$\hat{\cal O} = ({\bf j}_V, {\bf j}_A, f_{\pi}\hat\sigma , \hat\pi )$
may be defined as
\be
\hat{\cal O} (x) = \frac{\delta\hat{\bf I}}{\delta\hat{\phi}} =
-i\hat{\cal S}^{\dagger} \frac{\delta \hat{\cal S}}{\delta\hat{\phi}}
\,\,\,.
\label{A15}
\ee
The chain rule yields
\be
&&{\bf V}_{\mu}^a (x) = {\bf j}_{V\mu}^a (x) + f_{\pi} {\b a}_{\mu}^{ac} (x)
\hat{\pi}^c (x)\nonumber\\&&
{\bf A}_{\mu}^a (x) = {\bf j}_{A\mu}^a (x) + f_{\pi}^2 a_{\mu}^a (x)
-f_{\pi} (\nabla_{\mu}\hat\pi )^a (x) \nonumber\\&&
\sigma (x ) =\hat{\sigma } (x) + f_{\pi}\nonumber\\&&
\pi^a(x) = \hat{\pi }^a (x)
\label{A16}
\ee
Substitution into (\ref{A3}) gives
\be
\nabla^{\mu} {\bf j}_{V\mu} + {\b a}^{\mu} {\bf j}_{A\mu } + {\b J} \pi =0
\label{A17}
\ee
and therefore
\be
\bigg( {\bf X}_V +{\b J} \frac {\delta}{\delta J} \bigg) {\hat {\cal S}} =0
\,\,\,.
\label{A18}
\ee

On the other hand, substitution into (\ref{A4}) gives
\be
&&\nabla^{\mu} {\bf j}_{A\mu} + {\b a}^{\mu} {\bf j}_{V\mu} =\nonumber\\&&
-f_{\pi}^2 \nabla^{\mu}a_{\mu} + f_{\pi} \nabla^{\mu}\nabla_{\mu} \pi
\nonumber\\&&-f_{\pi}
{\b a}^{\mu}{\b a}_{\mu} \pi +
f_{\pi} (m_{\pi}^2 + s ) \pi \nonumber\\&&
- (J-f_{\pi}\nabla^{\mu} a_{\mu} )
(\hat\sigma +f_{\pi} )\,\,\,.
\label{A19}
\ee
This equation may be integrated by introducing the retarded and advanced
Green's functions
\be
\bigg( -\Box -m_{\pi}^2 -{\bf K} \bigg) \,G_{R,A} = {\bf 1}
\label{A20}
\ee
\be
{\bf K} = 2{\b v}^{\mu}\partial_{\mu} + (\partial^{\mu}{\b v}_{\mu}) +
{\b v}^{\mu}{\b v}_{\mu} -{\b a}^{\mu}{\b a}_{\mu} +s
\label{A21}
\ee
where we have adopted a condensed matrix notation. We  have the
Yang-Feldman-Kallen type-equations \cite{yang}
\be
\pi =&&\bigg( 1 +G_R {\bf K} \bigg) \pi_{\rm in} - G_R J
+G_R \bigg( \nabla^{\mu} a_{\mu} - J/f_{\pi} \bigg) \hat\sigma
\nonumber\\&&
-\frac 1{f_{\pi}} G_R \bigg( \nabla^{\mu}{\bf j}_{A \mu} +{\b a}^{\mu}
{\bf j}_{V\mu }\bigg) \nonumber\\ =&&
\bigg( 1 +G_A {\bf K} \bigg) \pi_{\rm out} - G_A J
+G_A \bigg( \nabla^{\mu} a_{\mu}- J/f_{\pi} \bigg) \hat\sigma
\nonumber\\&&
-\frac 1{f_{\pi}} G_A \bigg( \nabla^{\mu}{\bf j}_{A \mu} +{\b a}^{\mu}
{\bf j}_{V\mu }\bigg)\,\,\,.
\label{A22}
\ee
Noting that $\pi_{\rm out} = {\hat{\cal S}}^{\dagger} \pi_{\rm in} \hat{\cal
S}$, and using (\ref{A15}) we arrive at
\be
\frac{\delta}{\delta J} \hat{\cal S} =&& -iG_R J \hat{\cal S} +
i\hat{\cal S} \bigg( 1+ G_R {\bf K}\bigg) \pi_{\rm in} \nonumber\\&& +\frac
1{f_{\pi}}
G_R \bigg( \nabla^{\mu} a_{\mu} - J/f_{\pi} \bigg) \frac{\delta \hat{\cal
S}}{\delta
s}-\frac 1{f_{\pi}}
G_R {\bf X}_A \hat{\cal S}\nonumber\\ =&&
-i G_A J \hat{\cal S} +
i  \bigg( 1+ G_A {\bf K}\bigg) \pi_{\rm in} \hat{\cal S}\nonumber\\&& +
\frac 1{f_{\pi}}
G_A \bigg( \nabla^{\mu} a_{\mu} - J/f_{\pi} \bigg)
\frac{\delta \hat{\cal S}}{\delta
s}-\frac 1{f_{\pi}}G_A {\bf X}_A \hat{\cal S}\,\,\,.
\label{A23}
\ee

Evidently, any result which is a consequence of (\ref{A12}) and symmetry
(\ref{A7}-\ref{A8}) must be contained in (\ref{A18},\ref{A23}).
Since (\ref{A18}) simply represents local isospin invariance, the nontrivial
results of current algebra must be basically contained in (\ref{A23}).

To show that this is the case and that (\ref{A23}) is the desired master
formula, we note that
\be
G_{R,A} =&& \Delta_{R, A} + \Delta_{R, A} {\bf K} G_{R, A} \nonumber\\=&&
\Delta_{R, A} + G_{R, A} {\bf K} \Delta_{R, A}
\label{A24}
\ee
where ${\Delta}_{R, A}$ are the Green's functions for free fields.
Multiplying (\ref{A23}) by
$(1+ G_A {\bf K})^{-1} = 1-\Delta_A {\bf K}$ and Fourier decomposing yield
\be
\bigg[ a_{\rm in}^a (k) , \hat{\cal S} \bigg] = &&
 \int d^4y d^4z e^{ik\cdot y}
\bigg( 1 + {\bf K} G_R \bigg)^{ac} (y, z)
\nonumber\\&&\times
\bigg( -i \hat{\cal S} ({\bf K} \pi_{\rm in})^c (z)
+i \hat{\cal S} J^c (z)
\nonumber\\&&
-\frac 1{f_{\pi}}
\bigg(\nabla^{\mu}a_{\mu} - J/f_{\pi}\bigg)^c (z)
\frac{\delta\hat{\cal S}}{\delta s (z)} \nonumber\\&&
+\frac 1{f_{\pi}}
{\bf X}^c_A(z) \hat{\cal S}\bigg)
\nonumber\\
\label{A25}
\ee
\be
\bigg[\hat{\cal S} ,  a_{\rm in}^{a\,\dagger} (k) \bigg] = &&
 \int d^4y d^4z e^{-ik\cdot y}
\bigg( 1 + {\bf K} G_R \bigg)^{ac} (y, z)
\nonumber\\&&\times
\bigg( -i \hat{\cal S} ({\bf K} \pi_{\rm in})^c (z)
+i \hat{\cal S} J^c (z) \hat{\cal S}
\nonumber\\&&
-\frac 1{f_{\pi}}
\bigg(\nabla^{\mu}a_{\mu} - J/f_{\pi}\bigg)^c (z)
\frac{\delta\hat{\cal S}}{\delta s (z)} \nonumber\\&&
+\frac 1{f_{\pi}}
{\bf X}^c_A(z) \hat{\cal S}\bigg)
\nonumber\\
\label{A26}
\ee
where $a_{\rm in}^a (k)$ and $a_{\rm in}^{a\dagger} (k)$
are the annihilation and
creation operators of incoming pions with momentum $k$ and isospin $a$.
Iterations give the two and  higher pion reduction formulas, $e.g.$
to order ${\cal O} (\phi)$
\be
&&\bigg[ a_{\rm in}^b (k_2) , \bigg[ \hat{\cal S} , a_{\rm in}^{a\dagger} (k_1)
\bigg]\bigg] =\nonumber\\&&
\int d^4y e^{-ik_1\cdot y} \frac 1{f_{\pi}}{\bf X}_A^a (y)
\bigg[ a_{\rm in}^b (k_2) , \hat{\cal{S}} \bigg] \,.
\label{A266}
\ee

The Bogoliubov causality condition \cite{bogo} implies that
\be
T^*\bigg( \hat{\cal O} (x_1) ....\hat{\cal O} (x_n) \bigg)
=(-i)^n {\hat{\cal S}}^{\dagger}
\frac{\delta^n }{\delta\hat{\phi} (x_1 ) ...\delta\hat{\phi} (x_n) }
\hat{\cal S}\,\,.
\label{A27}
\ee
With this in mind, using (\ref{A25}-\ref{A266}),
sandwiching between nucleon states and
switching off the external fields, give the familiar $\pi N$ scattering
formula
\be
&&
<N(p_2 ) | \bigg[a_{\rm in}^b (k_2) ,\bigg[ {\bf S} , a_{\rm
in}^{a\dagger}(k_1)\bigg]\bigg] |N(p_1 ) > =\nonumber\\&&
-\frac i{f_{\pi}} m_{\pi}^2 \delta^{ab} \int d^4y e^{-i(k_1-k_2)\cdot y}
<N(p_2 ) | {\hat \sigma} (y ) | N(p_1) > \nonumber\\&&
-\frac 1{f_{\pi}^2} k_1^{\alpha} k_2^{\beta} \int d^4y_1 d^4 y_2 e^{-ik_1\cdot
y_1 + ik_2\cdot y_1} \nonumber\\&&
\qquad\times <N(p_2) | T^*\bigg(
{\bf j}_{A\alpha}^a (y_1) {\bf j}_{A\beta}^b (y_2 ) \bigg) | N(p_1)
>\nonumber\\&&
+\frac 1{f_{\pi}^2} k_1^{\alpha} \int d^4y e^{-i (k_1-k_2)\cdot y}
\epsilon^{abe} <N(p_2) | {\bf V}_{\alpha}^e (y) | N(p_1) >
\nonumber\\
\label{A28}
\ee
where ${\bf S} = {\hat{\cal
S}}|_{\phi =0}$ is the on-shell S-matrix. The disconnected part in (\ref{A28})
can be checked to cancel. At threshold, (\ref{A28}) yields
the Tomozawa-Weinberg relation \cite{tomozawa}.

The extension to $\pi\pi$ scattering is straightforward in principle, although
lengthy in practice. We find that the transition amplitude
$i{\cal T} ( p_2 d, k_2 b \leftarrow k_1 a, p_1 c )$
is a sum of four contributions
\be
i{\cal T}_{\rm tree} = \frac i{f_{\pi}^2} \bigg( s-m_{\pi}^2 \bigg)
\,\,\delta^{ac}\delta^{bd} +\,\,{\rm 2\,\, perm.}
\label{A30}
\ee
\be
i{\cal T}_{\rm rho} =&& \frac i{f_{\pi}^2}\epsilon^{abe}\epsilon^{cde}
\bigg( {\bf F}_V (t) - 1 -\frac t{4f_{\pi}^2} {\bf \Pi}_V (t) \bigg)
\nonumber\\&&+\,\,{\rm {2\,\,perm.}}
\label{A31}
\ee
\be
i{\cal T}_{\rm sigma} =&& -\frac {2im_{\pi}^2}{f_{\pi}} \delta^{ab}\delta^{cd}
\bigg( {\bf F}_S (t) +\frac 1{f_{\pi}} -\frac 1{2f_{\pi}^2}
<0|\hat{\sigma}  |0> \bigg) \nonumber\\&&
+\frac {m_{\pi}^4}{f_{\pi}^2}  \delta^{ab} \delta^{cd} \int d^4 y e^{-i
(k_1-k_2)\cdot y}
\nonumber\\&&\times
<0| T^*\bigg( \hat\sigma (y) \hat\sigma (0) \bigg) |0>_{\rm conn.} +
\,\,{\rm {2\,\, perm.}}
\label{A32}
\ee
\be
i{\cal T}_{\rm rest} =&&
+\frac 1{f_{\pi}^4} k_1^{\alpha} k_2^{\beta} p_1^{\gamma} p_2^{\delta}
\nonumber\\&&\times
\int d^4 y_1 d^4 y_2 d^4 y_3
\,e^{-ik_1\cdot y_1 +ik_2\cdot y_2 -ip_1\cdot y_3}\nonumber\\&&
<0| T^*\bigg( {\bf j}_{A\alpha}^a (y_1) {\bf j}_{A\beta}^b (y_2)
{\bf j}_{A\gamma}^c (y_3) {\bf j}_{A\delta}^d (0)\bigg) |0>_{\rm conn.}
\nonumber\\
\label{A33}
\ee
where $s, t, u$ are the Mandelstam variables,
\be
&&<0| a_{\rm in}^d (p_2) {\bf V}_{\alpha}^e (y) a_{\rm in}^{c\dagger  } (p_1)
|0>_{\rm conn.} =\nonumber\\&&
i\epsilon^{dec} (p_1 +p_2)_{\alpha} \,\,{\bf F}_V (t) e^{-i (p_1-p_2)\cdot y}
\label{A34}
\ee
is the pion electromagnetic form factor,
\be
&&i\int d^4x e^{iq\cdot x}
<0| T^*\bigg( {\bf V}_{\alpha}^a (x) {\bf V}_{\beta}^b (0) \bigg) |0> =
\nonumber\\&&\qquad
\delta^{ab} \bigg( -g_{\alpha \beta} q^2 + q_{\alpha} q_{\beta} \bigg)
{\bf \Pi}_V (q^2)
\label{A35}
\ee
is the isovector correlation function, and
\be
&&<0| a_{\rm in}^d (p_2) \sigma (y) a_{\rm in}^{c\, \dagger  }(p_1)
|0>_{\rm conn.} =\nonumber\\&&\qquad
\delta^{cd} \,\,{\bf F}_S (t) e^{-i (p_1-p_2)\cdot y}
\label{A36}
\ee
is the scalar form factor.
Experimentally, (\ref{A34}-\ref{A35}) are well described by $\rho$ dominance.

The unknown terms (\ref{A32}-\ref{A33}) may be estimated at low energies by
expanding in $1/f_{\pi}$. The master equation (\ref{A23}) then truncates to
\be
\frac{\delta{\hat{\cal S}_0}}{\delta J} =&& -i\hat{\cal S}_0
G_R J + i\hat{\cal S}_0 \bigg( 1 + G_R {\bf K} \bigg) \pi_{\rm in}\nonumber\\
=&&-i\hat{\cal S}_0
G_A J + i\bigg( 1 + G_A {\bf K} \bigg) \pi_{\rm in} \hat{\cal S}_0
\label{A37}
\ee
corresponding to the quadratic action
\be
{\bf I}_Q = &&\frac 12 \int d^4x \bigg(
(\nabla^{\mu} \pi )^a (\nabla_{\mu} \pi )^a -
({\b a}^{\mu} \pi )^a ({\b a}_{\mu} \pi )^a\nonumber\\&&\qquad\qquad -
(m_{\pi}^2 + s ) \pi^a \pi^a \bigg) +
\int d^4 x \,\, J^a \pi^a \,\,.
\label{A38}
\ee
In (\ref{A37}-\ref{A38}), $s$ and $a_{\mu}^a$ enter only through the
combination $\hat s = s{\bf 1} -{\b a}_{\mu}{\b a}^{\mu}$. If we take this to
be true for $\hat{\cal S}_0$, we obtain a two-parameter fit to pionic data at
one-loop level, which reproduces the KSFR relation
\cite{ksfr}. Also, since $\hat s$ is isospin
symmetric, the bulk of the $\rho$ contribution to $\pi\pi$ scattering at low
energies is given by (\ref{A31}) in a model independent manner.

With (\ref{A37}-\ref{A38}) and the assumption above, the sum
(\ref{A32}-\ref{A33}) is given by
\be
&&+\frac i{f_{\pi}^4} m_{\pi}^2 \delta^{ab}\delta^{cd} \bigg( 2t -\frac 52
m_{\pi}^2 \bigg) \bigg( \hat{c}_1 +{\cal J} (t) \bigg)\nonumber\\&&
+\frac i{4f_{\pi}^4}
\bigg(2\delta^{ab}\delta^{cd} + \delta^{ac}\delta^{bd} +
\delta^{ad}\delta^{bc}\bigg)
\nonumber\\&&\qquad\times \bigg(t-2m_{\pi}^2\bigg)^2
\bigg( \hat{c}_1 +{\cal J} (t) \bigg)\nonumber\\&&
+\,\,{\rm 2\,\,perm.}
\label{B1}
\ee
whereas (\ref{A34}-\ref{A35}) become
\be
{\bf F}_V (t) = 1 + \frac 1{2f_{\pi}^2}
\bigg( c_1t + \frac {t}{72\pi^2} + \frac 13 (t-4m_{\pi}^2) {\cal J} (t)\bigg)
\label{B2}
\ee
\be
\Pi_V (t) = c_1 + \frac 1{72\pi^2} + \frac 13 \bigg( 1-\frac
{4m_{\pi}^2}t\bigg) {\cal J} (t)
\label{B3}
\ee
where $c_1$ and $\hat{c}_1$ are the two constants and
\be
{\cal J} (q^2) = && -i \int \frac{d^4k}{(2\pi)^4}
\bigg( \frac 1{k^2-m_{\pi}^2 +i0}\frac 1{(k-q)^2-m_{\pi}^2 +i0}
\nonumber\\&&\qquad\qquad\qquad -
\bigg(\frac 1{k^2-m_{\pi}^2 +i0}\bigg)^2 \bigg) \nonumber\\
=&& \frac 1{16\pi^2} \int_0^1 dx \frac {x(1-2x) q^2}{x(1-x)q^2 -m_{\pi}^2 +i0}
\,\,\,.
\label{B4}
\ee
The $\rho$ data gives $c_1=0.035$ whereas a fit to the $\pi\pi$ scattering data
\cite{data} leads to seven determinations of $\hat{c}_1$
\be
16\pi^2 \hat{c}_1 =&& \frac{1024\pi^3}{63}\frac{f_{\pi}^4}{m_{\pi}^4}
\bigg( a_0^0 ({\rm exp}) -a_0^0 ({\rm tree})\bigg) -\frac {14}{9} \nonumber\\
=&&8\pm 5
\nonumber
\ee
\be
16\pi^2 \hat{c}_1 =&& \frac{64\pi^3}{9}\frac{f_{\pi}^4}{m_{\pi}^2}
\bigg( b_0^0 ({\rm exp}) -b_0^0 ({\rm tree}) -b_0^0 (\rm rho )\bigg)
\nonumber\\&&-\frac {91}{108}\nonumber\\ =&& 3\pm 1
\nonumber
\ee
\be
16\pi^2 \hat{c}_1 =&& 320\pi^3\, {f_{\pi}^4}
\bigg( a_2^0 ({\rm exp}) -a_2^0 (\rm rho )\bigg) +\frac {73}{180}
\nonumber\\=&&
2\pm 1
\nonumber
\ee
\be
16\pi^2 \hat{c}_1 = && 384\pi^3\, \frac{f_{\pi}^4}{m_{\pi}^2}
\bigg( a_1^1 ({\rm exp}) -a_1^1 ({\rm tree}) -a_1^1 (\rm rho )\bigg)
\nonumber\\&&
+\frac 14 \nonumber\\=&&
2\pm 5 \nonumber
\ee
\be
16\pi^2 \hat{c}_1 =&& \frac{512\pi^3}{3}\frac{f_{\pi}^4}{m_{\pi}^4}
\bigg( a_0^2 ({\rm exp}) -a_0^2 ({\rm tree})\bigg) -\frac 43 \nonumber\\=&&
26\pm 21 \nonumber
\ee
\be
16\pi^2 \hat{c}_1 = && \frac{128\pi^3}3 \frac{f_{\pi}^4}{ m_{\pi}^2}
\bigg( b_0^2 ({\rm exp}) -b_0^2 ({\rm tree}) -b_0^2 (\rm rho )\bigg)
\nonumber\\&&
-\frac {35}{36}\nonumber\\ =&&
-1\pm 2 \nonumber
\ee
\be
16\pi^2 \hat{c}_1 = && 640\pi^3\, {f_{\pi}^4}
\bigg( a_2^2 ({\rm exp}) -a_2^2 (\rm rho)\bigg) +\frac {19}{90} \nonumber\\=&&
3\pm 1 \nonumber\\
\label{B5}
\ee
which is seen to be consistent, to the possible exception of
$16\pi^2\hat{c}_1 = -1\pm 2$. Here $a_l^I$ and $b_l^I$ stand respectively
for the scattering lengths and range parameters with isospin $I$ and
orbital momentum $l$.

A comprehensive discussion of the present formulation,
further applications and detailed
comparison with previous work by other authors will be given elsewhere
\cite{YZ2}. Extension to $SU_L(3)\times SU_R(3)$ is currently under
investigation.

\vglue 0.6cm
{\bf \noindent  Acknowledgements \hfil}
\vglue 0.4cm
This work was supported in part  by the US DOE grant DE-FG-88ER40388.

\vskip 1cm
\setlength{\baselineskip}{15pt}

\end{document}